\documentclass{pasj00}
\draft

\begin{document}
\SetRunningHead{Awaki et al.}{Detection of NGC 2273 with Suzaku}
\Received{**/**/**}
\Accepted{**/**/**}

\title{Detection of Hard X-Rays from the Compton-Thick Seyfert 2 Galaxy NGC 2273 with Suzaku}


\author{Hisamitsu \textsc{Awaki},\altaffilmark{1},
        Yuichi \textsc{Terashima}, \altaffilmark{1}, 
        Yuusuke \textsc{Higaki}, \altaffilmark{1}, 
              Yasushi \textsc{Fukazawa},\altaffilmark{2} 
        }
\altaffiltext{1}{Department of Physics, Ehime University, Matsuyama, 790-8577, Japan}
\altaffiltext{2}{Department of Physical Science, Hiroshima University, \\ 
                   1-3-1 Kagamiyama, Higashi-Hiroshima, Hiroshima 739-8526}
\email{awaki@astro.phys.sci.ehime-u.ac.jp}

%


\maketitle

\begin{abstract}
We have obtained a broad-band spectrum of the Compton-thick Seyfert 2
galaxy NGC 2273 with Suzaku. The spectrum reveals the first detection
of hard X-rays above $\sim$10 keV from NGC 2273. The broad-band
spectrum is well represented by a three-component model, accompanied
by both a strong iron K$\alpha$ line with an equivalent width of
$\sim$1.8 keV and several weak lines.  The three-component model
consists of a soft component, a reflection component from cold matter,
and an absorbed power-law component.  The soft component can be 
represented by thin thermal emission with $kT\sim$0.56 keV or by a
scattered component with a scattering fraction of 0.4\%.  Fixing the
photon indices of the power law and reflection components at 1.9, we
found that the power law component, heavily absorbed by gas with a
column density of $\sim$ 1.5$\times$10$^{24}$ cm$^{-2}$, has an
intrinsic 2--10 keV luminosity of $\sim$1.7$\times$10$^{42}$ erg
s$^{-1}$.  We also apply a reflection model based on a Monte Carlo
simulation, assuming a simple torus geometry.  We found that the model
fits the broad band spectrum well, and we  place some tentative 
constraints on the geometry of the putative torus in NGC2273.
\end{abstract}

\section{Introduction}

Compton-thick active galactic nuclei (AGN) are predicted to constitute
about a half of AGN (e.g., \cite{risaliti99}). This type of AGN has
a characteristic strong iron line with an equivalent width of $>$ 1
keV (e.g., \cite{bassani99}), and a low ratio of $L_{\rm X}$/$L_{\rm
  [OIII]}$ (e.g., \cite{maiolino98}), where $L_{\rm X}$ and $L_{\rm
  [OIII]}$ are the luminosity in the 2--10 keV band and the luminosity
of the [OIII]$\lambda$5007 line, respectively.  However, the detailed
nature of this type of AGN remains unclear due to the heavy
obscuration below 10 keV and the complex X-ray spectrum.  To reveal the
nature of the hidden nucleus and to understand the complex X-ray
spectra, we need to obtain broad-band spectra covering energies above
10 keV.  BeppoSAX detected hard X-rays above 10 keV from several
Compton-thick Seyfert 2 galaxies: e.g., NGC 1068 (Matt et al. 1997),
NGC 4945 (Guainazzi et al. 2000), Mrk 3 (Cappi et al. 1999), Circinus
galaxy (Matt et al. 1999), NGC 3393, and NGC 4939 (Maiolino et
al. 1998).  Thanks to observations above 10 keV with INTEGRAL, 
Swift/BAT, and Suzaku, the number of detected Compton-thick
sources  has increased to 18 (Della Ceca 2008, and references
therein).  However, the number of Compton-thick sources whose
wide-band spectra were analyzed in detail is small:  e.g., NGC 1068
(Matt et al. 1997), Circinus galaxy (Matt et al. 1999; Yang et
al. 2008), Mrk 3 (Cappi et al. 1999; Awaki et al. 2008), NGC 4945
(Guainazzi et al. 2000; Itoh et al. 2008), and ESO 005-G004 (Ueda et
al. 2007). 
   
NGC 2273 is a nearby SB(r)a galaxy with a Seyfert 2 nucleus
(z=0.006138), and it is listed in a bright [OIII] sample of
\citet{risaliti99}. This galaxy is considered  to be a Compton-thick
Seyfert 2 galaxy due to the detection of a strong iron line with an 
equivalent width of greater than 1 keV and a low ratio of $L_{\rm
  X}$/$L_{\rm [OIII]}$ (e.g., \cite{guainazzi05}). The
existence of a hidden nucleus is also suggested by the detection of
polarized broad lines \citep{moran00}, establishing the presence of a
Seyfert 1 nucleus obscured by thick matter. To measure the absorbing
column, NGC 2273 was observed with ASCA, BeppoSAX, and 
XMM-Newton, but the column density could not be determined from these
data.  \citet{terashima02} found that NGC 2273 had a large column
density of 1.1$\times$10$^{24}$ cm$^{-2}$, by fitting the 0.8--10 keV
ASCA spectra with a partially-covered power-law model. On the other
hand, \citet{maiolino98} and \citet{guainazzi05} suggested a large
absorbing column $>$10$^{25}$ cm$^{-2}$, due to lack of the absorbed
power-law component in the spectra below 10 keV observed with BeppoSAX
and XMM-Newton.  In order to measure the column density and to reveal
its nuclear activity, a wide-band spectrum of NGC 2273 is crucial. 

Using the Japanese X-ray satellite Suzaku, it is possible to obtain a
wide-band spectrum from 0.2 to 700 keV \citep{mitsuda07}. The
satellite is suitable for achieving the aims of detecting the high
energy X-rays above 10 keV and of revealing the properties of the NGC
2273 nucleus.  In this paper, we present the analysis of the
broad-band spectrum of NGC 2273 obtained with Suzaku, and present the
results by fitting the spectrum with a baseline model consisting of a
soft component, an absorbed power-law component, and a reflection
component.  In addition, we  apply a reflection model based on a
Monte Carlo simulation assuming a simple torus geometry
\citep{ikeda08}. Quoted errors correspond to the 90 \% confidence
level for one interesting parameter, unless explicitly stated.

\section{Observations and Data Reduction}

NGC 2273 was observed with the Suzaku satellite on 2007 April 21--22
during the AO-2 phase.  Suzaku has four X-ray telescopes (XRTs:
\cite{serlemitsos07}) with X-ray CCD cameras (XIS) in their
focal-planes.  The XISs are sensitive to 0.2--12 keV X-rays on a
18$^{\prime}\times18^{\prime}$ field of view. Three of the four XISs
(XIS-0, 2, and 3) are front-illuminated CCDs (FI) and the other XIS
(XIS-1) is a back-illuminated CCD (BI) \citep{koyama07}. Due to 
damage to XIS2 on 2006 November 9, no astronomical data were obtained
with XIS2.  Suzaku also has a non-imaging hard X-ray detector (HXD:
\cite{takahashi07}).  The HXD has two types of  detectors, the PIN and
the GSO, which are sensitive between $\sim$10 and  $\sim$700 keV.

NGC 2273 was placed at the HXD-nominal position, and then both XIS
and HXD data were obtained simultaneously.  The XISs were operated in
the normal clocking mode with spaced-raw-charge injection (SCI) in
order to recover the cumulative effect of in-orbit radiation damage
\citep{nakajima08}.  The observed data were processed through pipeline
processing. We analyzed the v2.0 data by using the analysis software
package HEASOFT 6.4 and the calibration database CALDB released on
2008 February 1.

\subsection{XIS Reduction}
The XIS team recommends a recalculation of the PI values to account
for the update of the CALDB files. Thus, we performed the
recalculation of unfiltered event files, and then reprocessed the
cleaned event files from the XIS-0, 1, and 3 with the {\it xselect}
script provided by NASA/GSFC GOF \footnote{The Suzaku Data Reduction
  Guide January 2008, which can   be obtained from the GOF web page at
  {\it
    http://heasarc.gsfc.nasa.goc/docs/suzaku/aehp\_data\_analysis.html}}.  
We imposed an additional data screening with a criterion "COR $>$ 6",
where COR is geomagnetic cut-off rigidity in units of GV.  The net
exposure of the cleaned event data was about 76 ks.  We made a 0.4--2
keV band image and a 2--10 keV band image using the cleaned XIS data
(Figure 1), and clearly detected NGC 2273 and two nearby bright X-ray
sources, 2XMM J065012.9+604842 and 2XMM J065003.7+604639.  The XIS0
count rates of these detected sources are listed in Table 1.  The
former bright XMM-source is also known as a radio source NVSS
J065012+604842 \citep{condon98}.  Since one of the serendipitous X-ray 
sources, 2XMM J065012.9+604842 is within  $\sim$2$^{\prime}$ .1
of NGC 2273, we selected X-ray events within a 2$^{\prime}$.0
radius centered on NGC 2273 and removed X-ray events close to the
serendipitous source for the spectral analysis in order to reduce the 
contamination from the serendipitous X-ray source.

Response matrices (RMFs) and ancillary response files (ARFs) were
generated for each XIS independently by using {\it xisrmfgen} and {\it
  xissimarfgen} \citep{ishisaki07}.  To examine the accuracy of the
energy scale of our data, we extracted spectra of the $^{55}$Fe
calibration sources which illuminate the CCD corners, and
then fitted the spectra with a two-Gaussian model.  The difference
between the best-fit center energies of the calibration sources and
the weighted value of the theoretical energies for the transitions of
K$\alpha_1$ and K$\alpha_2$ was 0--10 eV, which was within the
accuracy ($\sim$0.2\% at Mn-K$\alpha$) of the energy calibration of
XIS.  The line widths of the calibration sources were found to be $<$
25 eV.  Since the energy scales of the front-illuminated CCDs, XIS-0
and 3 were quite similar to within 5 eV, we added the XIS-0 and XIS-3
spectra of NGC 2273 using {\it mathpha}.

To estimate the contamination from the nearby source, we made a
spectrum of 2XMM J065012.9+604842 by extracting X-ray events within a
2$^{\prime}$ radius centered on this source. Note that the X-ray events
close to NGC 2273 were removed in this spectral analysis. The X-ray
spectrum is shown in Figure 2.  The spectrum was fitted by a single
power law model with a photon index of 1.72$^{+0.15}_{-0.14}$ absorbed
by $N_{\rm H}$=7$^{+7}_{-6}\times$10$^{20}$ cm$^{-2}$. The absorbing
column is consistent with the Galactic column density toward NGC 2273,
6.7$\times$ 10$^{20}$ cm$^{-2}$ \citep{kalberla05}.  The X-ray fluxes
in the 0.5--2 keV and the 2--10 keV bands were measured to be
7.6$\times$10$^{-14}$ and 1.7$\times$ 10$^{-13}$ ergs s$^{-1}$
cm$^{-2}$, respectively.  We found, analyzing the XMM-Newton archival
data, that the flux had increased by a factor two in both bands.
The contamination of 2XMM J065012.9+604842 in the NGC 2273 source
region was inferred to be $\sim$1.8$\times$10$^{-14}$ erg s$^{-1}$
cm$^{-2}$ in the 2--10 keV band, and this contamination was taken into
account in our fitting procedure.

\subsection{HXD Reduction}
The internal (non-X-ray) background (NXB) of the HXD-PIN is variable,
and depends strongly on the Suzaku satellite orbit \citep{kokubun07}.
Thus we extracted both the source and the NXB spectra with identical
good time intervals.  After the instrument dead-time correction of the 
source spectrum, an exposure time of 64.6 ks was obtained.  A NXB
spectrum for the NGC 2273 observation was produced by the tuned NXB
event file for v2.0 data.  Since the HXD-PIN is not an imaging
detector, we also had to subtract any contribution from the Cosmic
X-ray Background (CXB).  We simulated a CXB spectrum observed by the
HXD-PIN, and then added the simulated CXB spectrum to the NXB
spectrum.  The HXD-PIN count rate of NGC 2273 was found to be about
0.020$\pm$0.002 cts s$^{-1}$ in the 15--40 keV band, which was about
7\% of the NXB of the HXD-PIN. Note that the errors presented in this
subsection are at 1$\sigma$ level.  The accuracy of the present PIN
NXB model in a 40 ks observation is about 1\% of the NXB at the
1$\sigma$ level (Mizuno et
al. 2008\footnote{http://www.astro.isas.jaxa.jp/suzaku/doc/suzakumemo/suzakumemo-2008-03.pdf}).
NGC 2273 was detected in the 15--40 keV band with $>$ 5 $\sigma$
level.  We confirmed the detection of the hard X-rays of NGC 2273 from
the comparison of the PIN count rates in the 15--40 keV band with
those obtained during earth occultation.  Although there were no
earth occultation data taken during this observation, we were able to
use the earth occultation data from both the previous and the
subsequent observations; these are included in the  Suzaku trend
archive provided by the Suzaku team. The upper panel  
of Figure 3 shows a light curve of the PIN count rates in the 15--40
keV band during earth occultation and the data reproduced by the NXB
model.  We also plotted the light curve of NGC 2273 in this panel. The
bottom panel shows the light curve in which the NXB model has been
subtracted. A clear excess was found during the NGC 2273 observation
with a mean count rate of  0.035$\pm$0.003 c s$^{-1}$. Subtracting the
CXB count rate of 0.016 c s$^{-1}$ from the mean count rate yields an
estimated net count rate of 0.019$\pm$0.003 c s$^{-1}$.  Since the PIN
FOV is 34$^{\prime}\times$34$\prime$ (Kokubun et al. 2007), there are bright
X-ray sources detected with XIS in the PIN FOV.  We estimated the
contamination of the brightest XMM source, 2XMM J065012.9+604842 by 
extrapolation of its 2-10 keV band spectrum. The contamination was
estimated to be about 0.0005 c s$^{-1}$ in the 15--40 keV band, which
corresponds to about 3 \% of the net count rate.  Since the 2--10 keV
count rate of the other XMM source is about 1/3 of the brightest XMM
source, we concluded that NGC 2273 was detected in the 15--40 keV
band.

The response matrix for a point source at the HXD nominal position
(ae\_hxd\_pinhxnome3\_20070914.rsp) was used for our spectral
analysis.  The cross-normalization of the HXD-PIN relative to the XIS0
have been derived using Suzaku observations of the Crab nebula. We
used the cross-normalization of 1.137$\pm$0.015 at the HXD nominal
position (Ishida et
al. 2007\footnote{http://www.astro.isas.jaxa.jp/suzaku/doc/suzakumemo/suzakumemo-2007-11.pdf}). 

The spectra of XIS and HXD PIN were binned to a minimum of 50 and 1500
counts per bin, respectively.

\section{Analysis \& Results}
\subsection{Spectral Analysis with the Baseline Model}

\citet{guainazzi05} found that the XMM-Newton spectrum  was well
fitted with a thin-thermal model ( {\it mekal} in 
{\it XSPEC}), plus Compton reflection model ({\it pexrav} ).  Thus we
fitted the XIS and HXD spectra with this model using {\it XSPEC}
v12.4.0. The abundances of \citet{ag89} were used.  The cosine of the
inclination and the iron abundance in {\it pexrav} were fixed at 0.45
and one solar abundance, respectively.  The best-fit spectrum and
residuals are shown in Figure 4.  The overall spectrum was represented
by this model with a $\chi^{2}$ of 199.2 with 130 degrees of freedom
(d.o.f.); however, there  were several features  apparent in the
residuals.  Therefore, we added gaussian components at $\sim$0.7,
$\sim$1.8, $\sim$3.2, $\sim$7.0, and $\sim$7.2 keV to the model. The
$\chi^{2}$ was reduced to 120 with 120 d.o.f.  The best-fit
parameters are listed as Model 1 in Table 2.  The $\sim$3.2 keV line
like feature was also detected in the BeppoSAX observation
\citep{maiolino98}, although the significance of this feature was low.
The best-fit photon index was well determined to be
1.49$^{+0.13}_{-0.08}$, which was consistent with that found in the
XMM-Newton spectrum (1.5$\pm$0.4; \cite{guainazzi05}).  The
thin-thermal component was represented by a temperature of
$kT\sim$0.59 keV and a metal abundance of 0.04 solar. We note that
a portion of the line fluxes of the additional lines at $\sim$0.7
and $\sim$1.8 keV could be explained by increasing metal abundances of
O and Si to 0.1$^{+0.4}_{-0.1}$ and 0.45$^{+0.30}_{-0.20}$ solar,
respectively. However, line-like features still remained near these
line energies. This may indicate that the lines are composed of
different origins.  The thermal component is absorbed by a thick
matter with $N_{\rm H}$=(2.9$^{+2.6}_{-1.4}$)$\times$10$^{21}$
cm$^{-2}$, which is significantly larger than the Galactic column
density.  The observed fluxes in the 0.5--2 keV, 2--10 keV, and 15--40
keV bands were estimated to be 6$\times$10$^{-14}$,
9.5$\times$10$^{-13}$, and 1.0$\times$10$^{-11}$ erg s$^{-1}$
cm$^{-2}$, respectively.  The 2--10 keV flux is nearly equal to the
BeppoSAX result \citep{maiolino98}, although the flux is 1.3 times
larger than that of XMM-Newton. The 0.5--2 keV flux is consistent with
the  XMM-Newton result considering the uncertainties
\citep{guainazzi05}.  The iron line intensity of
(2.5$\pm$0.2)$\times$10$^{-5}$ ph s$^{-1}$ cm$^{-2}$ is consistent
with that found using XMM-Newton
((2.3$^{+0.4}_{-0.3}$)$\times$10$^{-5}$ ph s$^{-1}$ cm$^{-2}$) and
using BeppoSAX ((2.36$^{+1.35}_{-0.70}$) $\times$10$^{-5}$ ph s$^{-1}$
cm$^{-2}$)). 

The best-fit photon index in the above model was clearly smaller than
the canonical value of $\sim$1.9 for AGNs (e.g., Nandra et 
al. 1997). Therefore, we added a heavily absorbed component to the
model.  We assumed that the photon index of the absorbed
power-law component is the same as that of the reflection component,
and fixed the photon index at 1.9.  The model gives a $\chi^{2}$ value
of 119  with 120 d.o.f.  The best-fit parameters are listed in Tables 2 and
3, and residuals are plotted in Figure 4 (Model 2 in Table 2).  Since
the best-fit absorption column ($N_{\rm H}$) was deduced to be
(1.46$\pm$0.42)$\times$10$^{24}$ cm$^{-2}$, we considered 
attenuation of the X-ray light due to Thomson scattering by the thick
matter along our line of sight.  The reflection ratio $R$ of the cold
reflection to the absorbed power law component was found to be 
0.5$^{+0.7}_{-0.3}$.  The 2--10 keV intrinsic luminosity, $L_{\rm
  2-10}$, was estimated to be 1.7$^{+2.5}_{-0.4} \times$10$^{42}$ erg
s$^{-1}$ assuming $H_{\rm 0}$=70 km s$^{-1}$ Mpc$^{-1}$.  The error of
the luminosity was calculated from the error on the normalization of
the absorbed power-law component.  The equivalent width (hereafter EW)
of the iron K$\alpha$ line with respect to the continuum emission was
deduced to be 1.8$\pm0.1$ keV.  The column density and the reflection
ratio are affected by the systematic uncertainties of the
reproducibility of the PIN NXB. For the +1\% NXB and $-$1\% NXB, the
best-fit values became 1.0 and 0.3 for $R$, and 1.15$\times$10$^{24}$
cm $^{-2}$ and 1.72$\times$10$^{24}$ cm$^{-2}$ for $N_{\rm H}$,
respectively. The change in these parameters is smaller than the
statistical errors on $R$ and $N_{\rm H}$.

$N_{\rm H}$ and $L_{\rm 2-10}$ are expected to depend on the
intrinsic power-law photon index.  We examined this parameter coupling
 by fitting the spectrum with model 2 assuming a photon index of 1.7 
 or 2.1 instead of 1.9. The models using these photon indices also fit
 well with $\chi^2$=116 and 124 (120 d.o.f), respectively. For a 
 photon index of 1.7, $N_{\rm H}$ and $L_{\rm 2-10}$ were found to be
 1.10$^{+0.55}_{-0.40}\times$10$^{24}$ cm$^{-2}$ and
 0.5$^{+1.0}_{-0.2}\times$10$^{42}$ erg s$^{-1}$, respectively.  On
 the other hand,  for a photon index of 2.1, $N_{\rm H}$ and $L_{\rm
   2-10}$ were found to be  1.6$\pm$0.4$\times$10$^{24}$ cm$^{-2}$ and
3.0$^{+1.7}_{-1.5}\times$10$^{42}$ erg s$^{-1}$, respectively.  These
best-fit values of $N_{\rm H}$ are close to those obtained with the
photon index was fixed at  1.9; in contrast, the $L_{\rm 2-10}$ values
are different from that using a photon index of 1.9.  Levenson et
al. (2006) found a relationship between Fe line luminosity $L_{\rm
  Fe}$ and intrinsic luminosity based on Monte Carlo simulations.
Although the ratio $L_{\rm Fe}$/$L_{\rm   2-10}$ depends on the
geometry of a torus, they found a typical ratio of 2$\times$10$^{-3}$.
NGC 2273 has a intense iron line with $L_{\rm Fe}$$\sim$
2$\times$10$^{40}$ erg s$^{-1}$, and the intense iron line requires
$L_{\rm 2-10}$$\sim$10$^{43}$ erg s$^{-1}$.  Based on the Levenson et
al.\ (2006) relationship,  the $L_{\rm  2-10}$ found using a  photon
index of 1.7 may be too low to explain the observed $L_{\rm Fe}$.

It is possible to search for a Compton shoulder (hereafter CS) on an
iron K$\alpha$ line with large equivalent width.  We added a pulse
function with a pulse width of 156 eV in order to represent the
first-order CS of iron K$\alpha$ line \citep{awaki08}. We found
that the fraction of the CS intensity with respect to that of the
primary Gaussian component is less than 15\%.  \citet{matt02}
calculated the fractions of CS for a spherical distribution
of  matter,  and for a plane-parallel slab.  The low fraction of $<$
15\% indicates a low column density below 10$^{24}$ cm$^{-2}$ and/or
small inclination angle to the slab.
 
An iron K$_{\beta}$ line was also detected with an intensity of
(3.2$\pm$1.0)$\times$10$^{-6}$ ph s$^{-1}$ cm$^{-2}$.  Fe K$_{\alpha}$
and K$_{\beta}$ lines are useful for estimating the ionization state
of the line-emitting gas (e.g., \cite{bianchi05}; \cite{yaqoob07}). We
derived a line intensity ratio, $I_{K\beta}$/$I_{K\alpha}$, of
0.13$\pm$0.04, which is consistent with emission by low ionization
states of iron \citep{palmeri03}. The measured center energies of the
iron K$\alpha$ and K$\beta$ lines also indicate a low ionization state
($\leqq$ Fe IX) of the reflecting matter.  The low ionization state 
corresponds to an ionization parameter ($\xi$=$\frac{L^{\rm
ion}}{n~r^{2}}$) of $\xi$ $<$ 1 erg cm s$^{-1}$, assuming low density
gas, where $L^{\rm ion}$, $n$ and $r$ are the luminosity in the 1 --
1000 ryd band, the hydrogen density of material, and the distance from
the X-ray source. Assuming a simple power law model with a photon
index of 1.9, $L^{\rm ion}$ is inferred to be 3.5-times the 2--10 keV
luminosity. 

The soft component of NGC 2273 was fitted with a thermal plasma
model with $kT\sim$0.56 keV. Next, we examined whether the soft component
instead could be represented by a scattered component as seen in Mrk
3. Since the soft component of Mrk 3 is a good template for the scattered
light of AGN (e.g., \cite{sako00}), we rescaled the soft component of
Mrk 3 to produce the soft component of NGC 2273.  In this fit, we
fixed the photon index at 1.8, which is the best-fit photon index for
Mrk 3 \citep{awaki08}.  We found that the overall spectrum of NGC 2273
could be represented with a scaling factor of $\sim$0.092.  The best-fit
parameters are listed as model 3 in Table 2, and the best-fit spectrum
is shown in Figure 5.  The scaling factor corresponds to a scattered
fraction of 0.4$\pm$0.1\%.  Note that we also included additional
lines that are present in the thin thermal model.  The best-fit
line intensities for the 0.7 keV and 1.8 keV lines are changed to
7$^{+7}_{-4}\times 10^{-6}$ and 1.2$\pm$10$^{-6}$ ph s$^{-1}$
cm$^{-2}$, respectively.

\subsection{Spectral Analysis with a Reflection Model based on a Monte
  Carlo Simulations} 

The baseline model is useful to reproduce X-ray spectra of Seyfert 2
galaxies empirically.  However, it is difficult to obtain information
about the structure of surrounding material from the spectral fitting
with the baseline model, since $pexrav$ was developed for an accretion
disc geometry (Magdziarz \& Zdziarski 1995).  Thus, we tried to
reproduce the X-ray spectrum of NGC 2273 with a reflection model based
on a Monte Carlo simulation in which a simple torus geometry of the
surrounding material with arbitrary half opening angle, viewing
angle, and column density along the equatorial plane was assumed (see
Figure 2 in \cite{ikeda08}). \citet{ikeda08} divided the simulation
spectra into three components: the direct component, a reflection
component absorbed by the torus itself (reflection 1), and an
unabsorbed reflection component (reflection 2), and created table
models of these reflection components for XSPEC.  We fitted the
observed spectrum with these table models in the energy range from 1
to 40 keV (model 4 in Table 4).  These models also worked well with a
$\chi^{2}$ of 108 (dof=106) (see Table 4 and Figure 6).  Due to 
Compton scattering by thick matter along our line of sight, the
absorbed component in Figure 6 was reduced by a factor of
$\exp(-\sigma_{\rm c}N_{\rm H})$, where $\sigma_{\rm c}$ is the Compton
scattering cross-section.  \citet{ikeda08} pointed out that the
$\chi^{2}$ value was affected by the statistical deviation of the
table models.  In the case of Mrk 3, the scatter in $\chi^{2}$ due
to that deviation was estimated to be about 5.  Since NGC 2273 is
fainter than Mrk 3, the scatter of the $\chi^{2}$ was estimated to be
less than 1. The best-fit opening angle of the torus was found to
be 40$^{+25}_{-30}$$^{\circ}$.  We found that the viewing angle was
strongly coupled to the opening angle, and the best-fit viewing
angle was 2--3$^{\circ}$ larger than the opening angle. In Table 4, we
give the difference between these angles when we fixed the opening
angle at the best-fit value.  This result means that we observe the
nucleus along a line of sight intercepting the torus near its edge. As
a result, we observed strong unabsorbed reflection. The best-fit
photon index of the power-law component is 2.42$^{+0.08}_{-0.57}$. The
positive bound on the photon index is a consequence of the hard limit
of the table models of $\Gamma=2.5$.  The column density along the
equatorial plane of the torus is 5.4$^{+3.7}_{-3.1}\times$ 10$^{24}$
cm$^{-2}$.  The X-ray luminosity in the 2--10 keV band was estimated
to be 3.9$\times$10$^{42}$ erg s$^{-1}$.  When we fixed the photon
index at 1.9, the $\chi^{2}$ becomes 111.6 (dof=107), the column
density along the equatorial plane becomes 
(4.8$^{+5.2}_{-3.0}$)$\times$10$^{24}$ cm$^{-2}$ and the intrinsic
X-ray luminosity in the 2--10 keV band is estimated to be
1.9$\times$10$^{42}$ erg s$^{-1}$.

\section{Discussion}
\subsection{Hard X-ray Emission and AGN Activity}
Thanks to the high sensitivity of Suzaku, we detected X-ray photons
above 10 keV at the $>$ 5 $\sigma$ significance level.  The hard X-ray
flux in the 15--40 keV band was estimated to be 1$\times$10$^{-11}$
erg s$^{-1}$ cm$^{-2}$.  The wide-band X-ray spectrum in the 0.3--40
keV band was roughly fitted with the model applied by
\citet{guainazzi05}. However, we found that the best-fit photon index
of $\sim$1.5 is clearly smaller than the canonical value for Seyfert
galaxies. Therefore, we added an absorbed power-law emission to the
model, and obtained the best-fit absorption column of
$\sim$1.5$\times$10$^{24}$ cm$^{-2}$, and the intrinsic luminosity of 
$L_{\rm 2-10}$=1.7$\times$10$^{42}$ erg s$^{-1}$.  Our observation has
revealed a luminous nucleus obscured by optically thick matter.

We compared the intrinsic X-ray luminosity of NGC 2273 with the
[OIII]$\lambda$5007 luminosity in order to determine the nature of the
Compton-thick nucleus.  \citet{heckman05} shows that ratios of hard
X-ray (2--10 keV) to [OIII] luminosities for Seyfert 1 galaxies are
distributed around 1.59 dex with a standard deviation of only 0.48
dex.  The [OIII] luminosity of NGC 2273 was found to be
4.2$\times$10$^{40}$ erg s$^{-1}$ by \citet{whittle92} assuming
$H_{\rm 0}$ = 70 km s$^{-1}$ Mpc$^{-1}$.  Based on the mean ratio
given by \citet{heckman05}, this [OIII] luminosity suggests that the
intrinsic X-ray luminosity from the AGN should be 1.6$\times$10$^{42}$
erg s$^{-1}$ in the 2--10 keV band.  This estimate is consistent with
our result.  Thus, the ionized region emitting the [OIII] line is most
likely powered by the intense emission from the nucleus.
\citet{maiolino98} pointed out that the Seyfert 2 galaxy NGC 
2273 was consistent with being Compton thick because of its low
luminosity ratio, $L_{\rm 2-10}$/$L_{\rm [OIII]}$, of $<$ 0 dex. Our
Suzaku observation allows us to measure the 2 - 10 keV intrinsic
luminosity of NGC 2273, and suggests that NGC 2273 has a similar
intrinsic ratio as those of Seyfert 1 galaxies.

We note that NGC 2273 was undetected in a 12 ks observation by the PDS
on-board BeppoSAX\citep{maiolino98}. They obtained a 1$\sigma$-upper
limit of 0.096 c s$^{-1}$ in the 15-100 keV band, which corresponds to
a 15-40 keV flux of $\sim$5$\times$10$^{-12}$ erg s$^{-1}$ cm$^{-2}$,
assuming the same spectral shape as that of Mrk 3.  This upper limit is
about half of our detected flux. This discrepancy can be explained by
time variability of the absorbed power-law emission on a time scale of
years.  If NGC 2273 lacked the absorbed power-law emission, the
observed flux would be reduced to $\sim$5$\times$10$^{-12}$ erg
s$^{-1}$ cm$^{-2}$.  The absence of the absorbed emission may be
caused by a decrease of the intrinsic luminosity and/or an increase of
the absorption column up to $>$ 3.5$\times$10$^{24}$ cm$^{-2}$.

\subsection{Soft X-ray Emission }
\citet{fwm00} found [NII] +H$\alpha$ emission associated with a
nuclear ring, and this emission was inferred to be produced by 
star-formation activity in the ring. The soft X-rays represented by a
thin-thermal model with $kT$$\sim$0.6 keV may be associated with the
star-formation activity.  The X-ray luminosity, $L_{\rm 0.5-2.0}$, of
the thin-thermal component was estimated to be 6$\times$10$^{39}$ erg
s$^{-1}$ in the 0.5 -- 2.0 keV band.  This galaxy has luminous
far-infrared emission with $L_{\rm FIR}$ = 2.9$\times$10$^{43}$ erg
s$^{-1}$ (Londsdale et al. 1992).  The observed 0.5--2 keV luminosity
is nearly equal to expectations given the ratio of $L_{\rm
0.5-2.0}$/$L_{\rm FIR}$ for starburst galaxies \citep{rcs02}.  In the
thin-thermal model, the absorption by $N_{\rm H}$=
(2.9$^{+2.6}_{-1.4}$)$\times$10$^{21}$ cm$^{-2}$ is required. The
absorption can be explained by an obscuration of HII regions in the
galaxy ring (Ferruit et al. 2000).  On the other hand, the soft X-rays
may come from a photoionized region as often seen in other Seyfert 2
galaxies.  The scattering fraction is found to be $\sim$0.4\%, which
is about half that found in Mrk 3 \citep{awaki08}.  In this model, the
additional lines at $\sim$0.7 keV and $\sim$1.8 keV were explained by
the emission from photo-ionized plasma. The line at $\sim$1.8 keV
suggests a lower-ionization form of Si in the photoionized region of
NGC 2273 compared with that of Mrk 3.

Since the observed soft emission is consistent both with thin thermal
emission from the star-forming region, and with a scattered component
often seen in Seyfert 2 galaxies,  we cannot conclusively determine
its origin.  A deep Chandra observation could possibly
separate the thin thermal emission from the nuclear emission, since
the H$\alpha$ region is often coincident with the thin-thermal
emission region, and the diameter of the circumnuclear region is about
4$^{\prime\prime}$. 

\subsection{Implications for reflecting matter in NGC 2273}
Using the iron lines, we found the reflecting matter is in a low
ionization state with $\xi$ $<$ 1 erg cm s$^{-1}$, suggesting that the
reflecting matter is located at some distance from the X-ray source:
$\sim 0.75 (\frac{L^{\rm ion}}{\rm 5\times10^{42}~erg~s^{-1}})^{1/2}~
(\frac{\xi}{\rm 1~erg~cm~s^{-1}})^{-1/2}~(\frac{n}{\rm
10^{6}~cm^{-3}})^{-1/2}$ (pc).  A characteristic broad-line region
size ($R_{\rm BLR}$) has been shown to be related to the X-ray, UV,
and optical continuum luminosities \citep{kaspi05}.  Using their
relationship, the characteristic $R_{\rm BLR}$ was estimated to be
$\sim 0.003 (\frac{L_{\rm 2-10}} {2\times10^{42}
  {\rm~erg~s^{-1}}})^{0.7}$ (pc). The distance of the reflecting
matter is larger than the characteristic size of broad line region of
NGC 2273.  It is natural to consider the reflection matter to be the
dusty torus which also obscures the broad line region in this Seyfert
2 galaxy.

We tried to determine the properties of the dusty torus of NGC 2273 by
fitting the reflection component seen in the observed spectrum with
the reflection model based on simulations.  Although the
opening angle of the torus was poorly constrained, we found that we
observe the nucleus along a line of sight intercepting the torus near
its edge.  If the photon index was fixed at 1.9, the intrinsic
luminosity in the 2--10 keV band and the column density of the torus
along the equatorial plane were estimated to be 1.9$\times$10$^{42}$
erg s$^{-1}$ and 4.8$^{+5.2}_{-3.0}\times$10$^{24}$ cm$^{-2}$,
respectively. The large column of the torus naturally explains the
observed large iron line EW of 1.8 keV (e.g., \cite{ikeda08}).
However, the large column predicts a large fraction of CS up to 0.2
(e.g., \cite{matt02}).  In the torus geometry, the column density
should be small, and/or the opening angle should be large to obtain
such a small CS.  We performed Monte Carlo simulations using the torus
geometry to explain the small CS. Figure 7 shows the fraction of the
CS as a function of column density and opening angle.  The large
opening angle seems to be inconsistent with the small fraction of the
scattering X-rays.  To resolve the puzzle about the small CS, an
accurate measurement of the CS using an instrument with high energy
resolution, such as the X-ray micro-calorimeter onboad {\it Astro-H},
is crucial. 

\section{Conclusion}
We observed the weak Compton-thick Seyfert 2 galaxy NGC 2273 with the
Japanese X-ray satellite Suzaku, and detected hard X-rays above 10
keV. The flux in the 15-40 keV is about 1$\times$10$^{-11}$ erg
s$^{-1}$ cm$^{-2}$.  First, the broad-band spectrum from 0.3 to 40 keV
was fitted with thin-thermal plus reflection components.  The best-fit
photon index was found to be 1.5, smaller than the
canonical value of Seyfert galaxies. We then added an absorbed
power-law component to the model.  When we fixed the photon index at
1.9, the absorbed column and intrinsic luminosity in the 2--10 keV
band were found to be 1.5$\times$10$^{24}$ cm$^{-2}$ and
2$\times$10$^{42}$ erg s$^{-1}$, respectively.  Our observation
reveals the obscured nucleus of NGC 2273. The luminosity ratio,
$L_{\rm X}/L_{\rm [OIII]}$, for NGC 2273 is similar to those for
Seyfert 1 galaxies. This indicates that the photoionized region that
is emitting the [OIII] lines is consistent with being powered by the
nuclear activity.  For the soft X-ray emission below 2 keV, it could
be represented equally well by thin-thermal emission or by a scattered
emission model as seen in Mrk 3, and thus we could not obtain a
conclusive result on the origin of the soft component.

A strong iron K$\alpha$ line with an EW of 1.8 keV was detected. To
achieve such a large EW, the direct light must be blocked by thick
matter with $N_{\rm H}$ $>$ 2$\times$10$^{24}$ cm$^{-2}$.  This column
density is consistent with the spectral fitting using a model based
on a Monte Carlo simulation. However, the large column density is
inconsistent with the observed small fraction of CS of the iron
K$\alpha$ line. We also detected a weak iron K$\beta$ line with a
center energy of $\sim$7.0 keV.  From the properties of K$\alpha$ and
K$\beta$ lines, we conclude that the iron emitting matter is in a low
ionization states of $<$ Fe VIII, which indicate that the iron
emitting matter is located far from the X-ray source.
  
We fitted the hard X-ray emission with a reflection model based on a
Monte Carlo simulation, assuming a simple torus geometry. Although the
opening angle of the torus was not constrained well, it was found that
we observe the nucleus along a line of sight intercepting the torus
near its edge. The column density of the torus along the equatorial
plane is 4.8$^{+6.2}_{-3.0}\times$10$^{24}$ cm$^{-2}$ if the photon
index is fixed at 1.9.

\vspace{1cm}

The authors wish to thank the members of the Suzaku team for their
operation of the satellite, and the members of the Suzaku HXD team for
their effort on the NXB study of the HXD.  We are very grateful to an
anonymous referee for valuable comments and to Dr. K. Leighly for her careful reading of our draft.  
This study is carried out in
part by the Grant support for Scientific Research of Ehime university
(H.A.)  and the Grant-in-Aid for Scientific Research (20740109 Y.T.)
of the Ministry of Education, Culture, Sports, Science and Technology.

\clearpage
\begin{table}
\begin{center}
\begin{scriptsize}
\caption{XIS0 count rates of the detected sources \label{tbl-1}}
\begin{tabular}{ccc}
\hline \hline
NAME & $N_{\rm 0.4-2 keV}^{*}$ & $N_{\rm 2-10keV}^{*}$ \\
   &  ($\times$10$^{-3}$ c s$^{-1}$) &    ($\times$10$^{-3}$ c s$^{-1}$)   \\  
\hline 
NGC 2273                              & 2.5$\pm$0.2  & 8.7$\pm$0.4 \\
2XMM J065012.9+604842 & 3.7$\pm$0.2 & 3.1$\pm$0.2  \\
2XMM J065003.7+604639 & 2.9$\pm$0.2 & 1.1$\pm$0.2  \\
\hline 
\end{tabular}

\end{scriptsize}
\end{center}
$*$ count rates within a 1$^{\prime}$ radius centered on each source \\
\end{table}

\begin{table}
\begin{center}
\begin{scriptsize}
\caption{Best-fit parameters of the three models to the wide-band spectrum of NGC 2273\label{tbl-1}}
\begin{tabular}{ccccccccc}
\hline \hline
ID & model$^{*}$ & $N_{\rm H1}$$^{\dagger}$ & soft component &  &    & hard component &    & $\chi^{2}$ (d.o.f.) \\
   &  &   & kT/scattering fraction & abundance &  photon index &  $N_{\rm H}$$^{\ddag}$ & $R^{\S}$ &  \\  
   &  &  ($\times$10$^{22}$ cm$^{-2}$)  & (keV/ --- ) &     &    &  ($\times$10$^{22}$ cm$^{-2}$) & & \\
\hline 
1 & T+R+lines                        & 0.29$^{+0.25}_{-0.15}$ &  0.59$^{+0.20}_{-0.14}$ & 0.04$^{+0.12}_{-0.03}$ &    
1.49$^{+0.13}_{-0.08}$ &  &  & 120 (120) \\
2 & T+R+abs-PL+lines  & 0.30$^{+0.28}_{-0.18}$ &   0.56$^{+0.14}_{-0.16}$ & 0.05$^{+0.17}_{-0.04}$ &   
1.9 (fixed) &  146$\pm42$  & 0.5$^{+0.7}_{-0.3}$ & 119 (120) \\
3 & S+R+abs-PL+lines & 0.14$\pm$0.07 &  0.092$^{+0.021}_{-0.018}$ & -  &     1.8 (fixed)   & 115$\pm40$  & 0.8$^{+1.2}_{-0.4}$ & 121 (122)  \\
\hline 
\end{tabular}

\end{scriptsize}
\end{center}
$*$ T: thin thermal plasma model ({\it mekal}). R: cold reflection model ({\it pexrav}). abs-PL: absorbed power-law emission. S: scattered emission.  \\
$\dagger$ the absorption column for each model, i.e. exp(-$\sigma$ $N_{\rm H1}$) (model). \\
$\ddag$ the absorption column for the absorbed power-law emission. \\
$\S$ the reflection ratio of the cold reflection to the absorbed power-law emission. \\
\end{table}


\begin{table}
\begin{center}
\begin{scriptsize}
\caption{The best-fit center energy and intensity of each line in model 2\label{tbl-2}}
\begin{tabular}{cccccccc}
\hline\hline 
Center energy$^{*}$ & line width  &  intensity & EW & line identifications \\
(keV)               &   (eV)  &  ($\times$10$^{-6}$ph s$^{-1}$ cm$^{-2}$)  & (eV) & \\
\hline 
0.745$^{+0.045}_{-0.025}$ &  5 (fixed) & 17$^{+100}_{-14}$  &  75$^{+440}_{-60}$   & OVII RRC ($>$ 0.739 keV)\\
1.806$^{+0.027}_{-0.025}$ & 10(fixed) & 1.7$\pm$0.7        & 160$\pm$70   & Si K$_{\alpha}$ (1.740 keV) and \\
                                 &             &                         &        & Si XIII (1.854(f)/1.865(r+i) keV)  \\
3.26$^{+0.06}_{-0.14}$       &  10(fixed) & 0.9$^{+0.7}_{-0.6}$ & 140 $^{+110}_{-100}$ &  Ar XVII (3.124(f)/3.140(r+i) keV) and \\ 
                                 &             &                         &          &  Ar XVIII (3.32 keV)  \\
6.400$\pm$0.005           & 4 ($<$26) & 24.7$\pm$1.7      &  1810$\pm$120  &  Fe K$_{\alpha}$  (6.4038 keV)\\
7.06$\pm0.03$              &  4$^{\dagger}$  & 3.2$\pm$1.0        &   270$\pm$85 & Fe K$_{\beta}$  (7.058 keV) \\
7.50$^{+0.09}_{-0.19}$     &       4$^{\dagger}$     & 1.4$^{+1.0}_{-1.1}$ &  130$^{+90}_{-100}$  &  Ni K$_{\alpha}$ (7.4782 keV)) \\
\hline 
\end{tabular}

\end{scriptsize}
\end{center}
$*$ corrected for  cosmological redshift ($z$=0.006138).\\
$\dagger$ these line widths were linked to that of the 6.4 keV line.
\end{table}

\begin{table}
\begin{center}
\begin{scriptsize}
\caption{Best-fit parameters of the cold reflection model based on a Monte Carlo simulation \label{tbl-3}}
\begin{tabular}{ccccccccccc}
\hline \hline
ID & model$^{*}$ & $N_{\rm H1}$ & soft component &  &    & hard component &  &   & $\chi^{2}$ (d.o.f.) \\
  &  &   & kT & abundance &  photon index &  $N_{\rm H}$$^{\dagger}$ & $\theta_{\rm oa}$$^{\dagger}$  & $\theta_{\rm i}$$^{\dagger}$-$\theta_{\rm oa}$   &  \\  
     &  &  ($\times$10$^{22}$ cm$^{-2}$)  & (keV) &     &    &  ($\times$10$^{22}$ cm$^{-2}$) & ($^{\circ}$) & ($^{\circ}$) &\\
\hline 
4 & T+RS+abs-PL+lines & 0.39 ($<$0.9) & 0.75$^{+0.25}_{-0.40}$ & 0.2 ($<$5) & 2.42$^{+0.08}_{-0.57}$ &  540$^{+370}_{-310}$ & 40$^{+25}_{-30}$ & 3$\pm$2$^{\ddag}$ &  108 (106) \\
\hline 
\end{tabular}

\end{scriptsize}
\end{center}
$*$ RS: cold reflection model based on a Monte Carlo simulation. \\
$\dagger$ $N_{\rm H}$: the column density along the equatorial plane of the torus. $\theta_{\rm oa}$: the opening angle of the torus. $\theta_{\rm i}$: the inclination angle of the torus. \\
$\ddag$ the best-fit value and error are listed, when $\theta_{\rm oa}$ is fixed at the best fit value.
\end{table}

\clearpage



\begin{figure}
\FigureFile(80mm,50mm){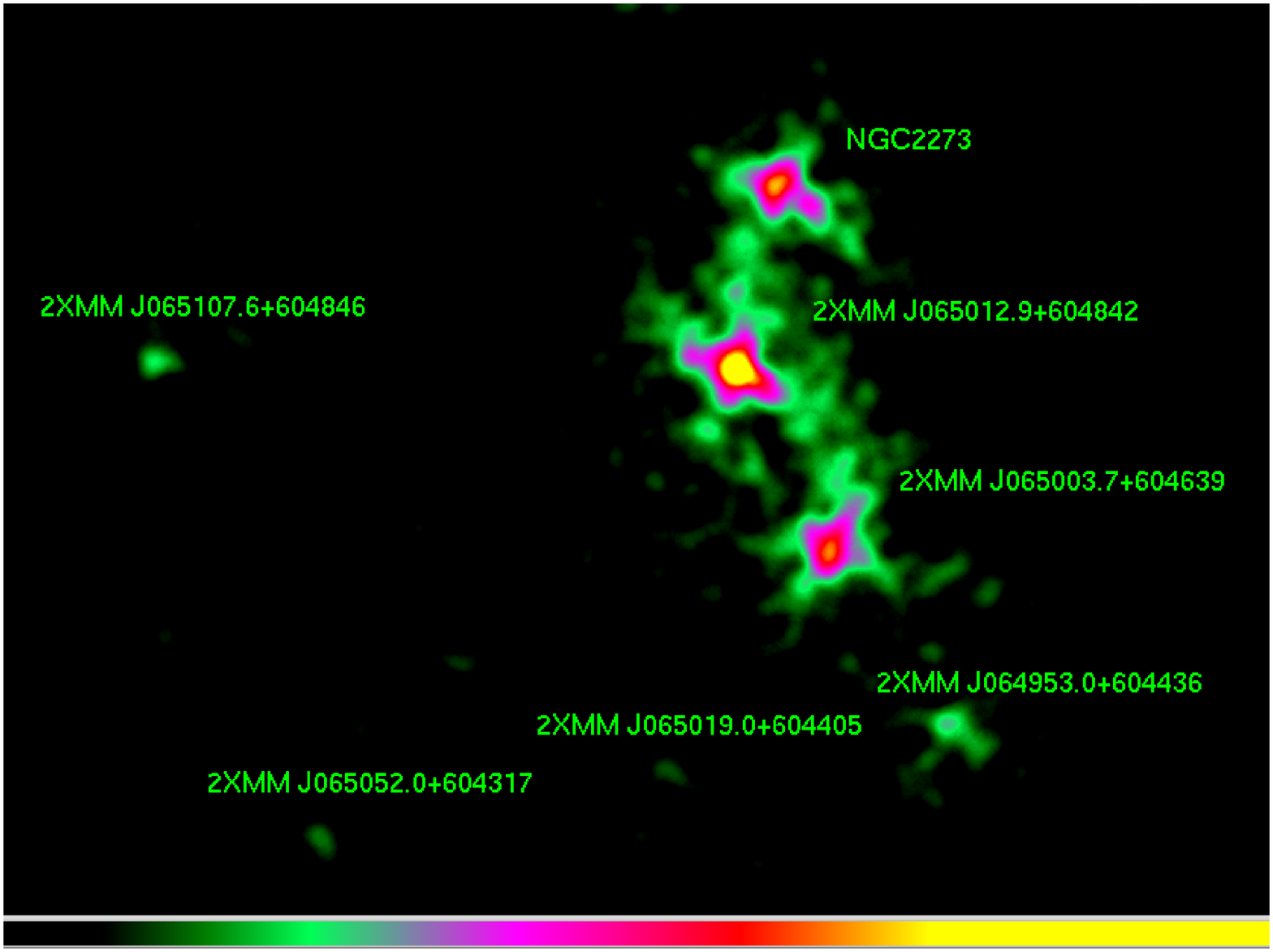}
\FigureFile(80mm,50mm){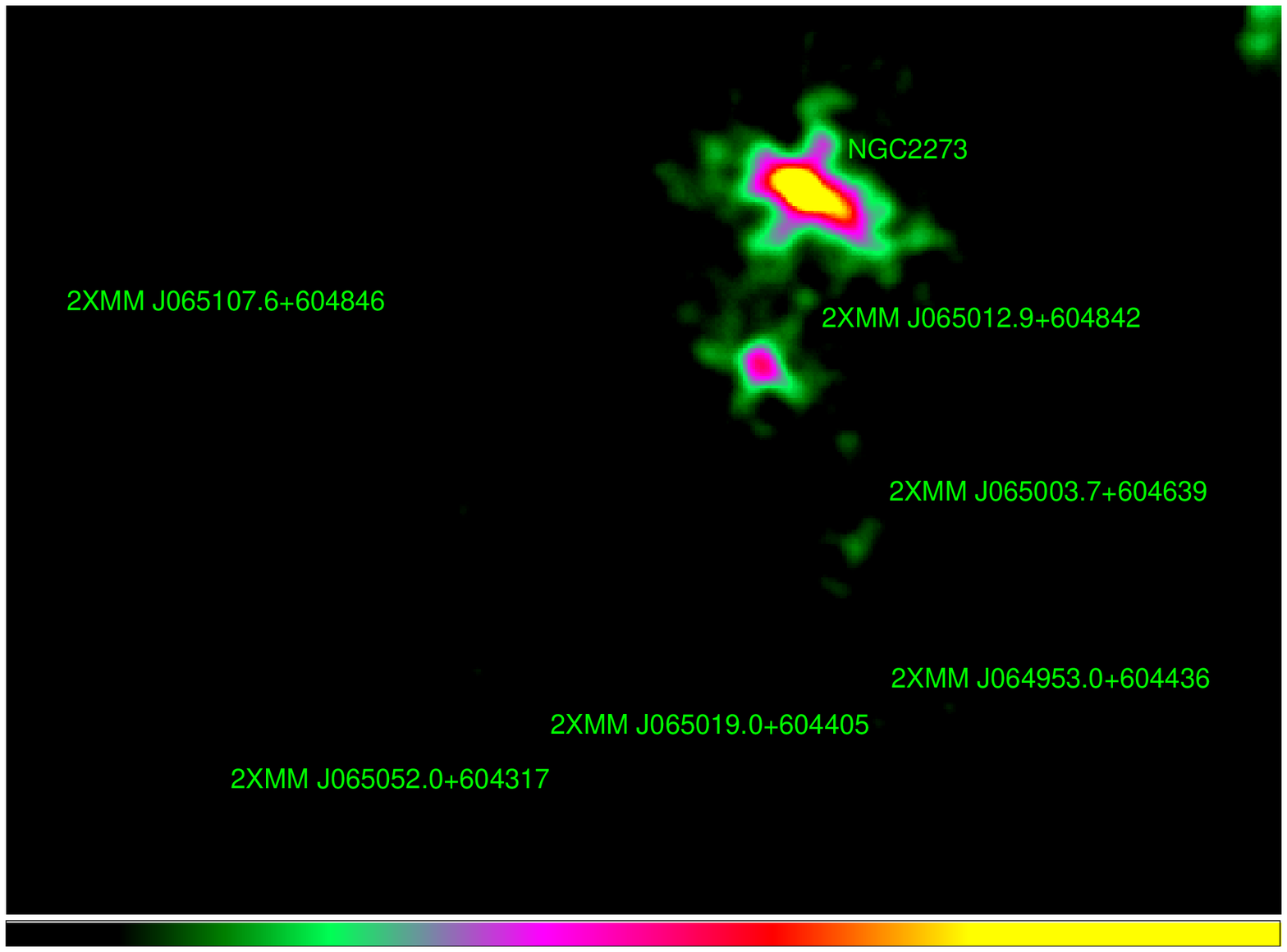}
\caption{The 0.4--2 keV (left) and 2--10 keV (right) images for NGC
2273 obtained by Suzaku.  The images were  smoothed by a gaussian 
function with $\sigma$=0$^{\prime}$.2.  The color scale is linear in
the range of 0.013--0.413 c pix$^{-1}$ for the 0.4--2 keV image and in
the range of 0.03--0.63 c pix$^{-1}$ for the 2--10 keV image. The
minimum value for each image corresponds to the background emission.  
 \label{fig1}}
\end{figure}

\begin{figure}
\includegraphics[scale=0.5, angle=-90, clip] {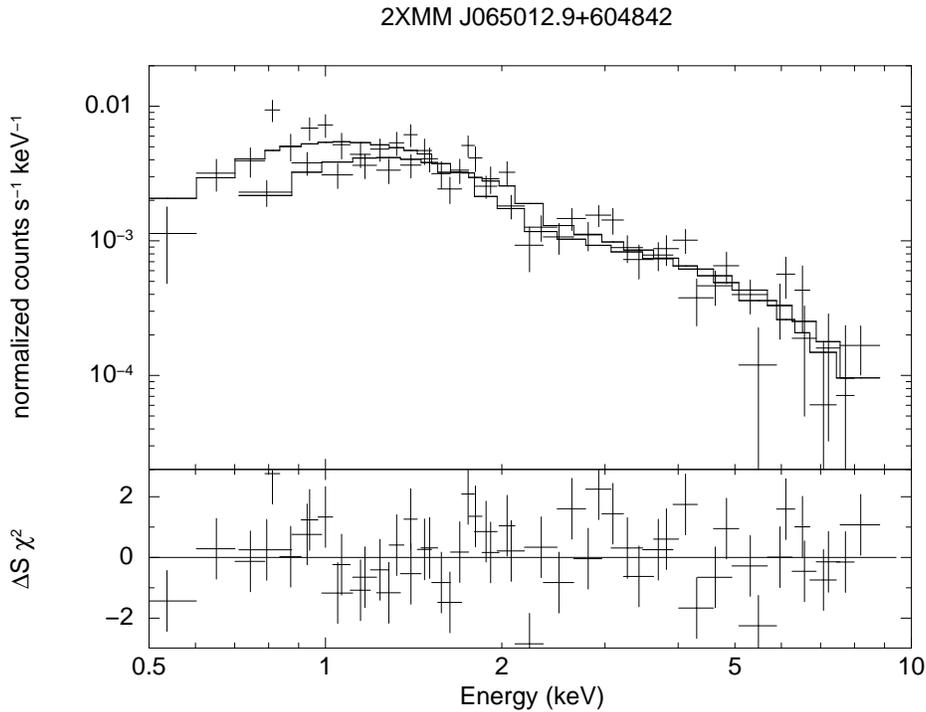}
\caption{The spectrum of 2XMM J065012.9+604842 observed with the
  Suzaku XIS. The spectrum is fitted with a  single power-law model.   
 \label{fig2}}
\end{figure}

\begin{figure}
\includegraphics[scale=0.5, angle=-90, clip] {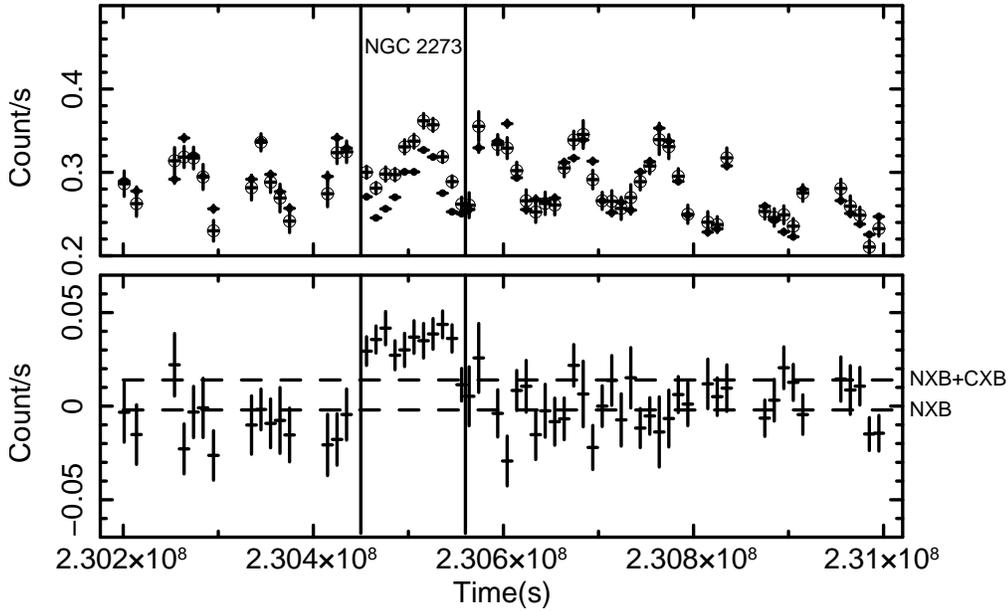}
\caption{The HXD PIN light curve in the 15--40 keV band. 
The upper panel shows observed count rates (open circles) and NXB
count rates reproduced from the tuned NXB events (filled circles). 
The PIN light curve consists of data from earth occultation periods
and the data from NGC 2273. The lower panel shows a light curve after
NXB subtraction. The "NXB" in the lower panel shows the mean count
rate of the earth occultation data after NXB subtraction. The CXB is the
count rate of the cosmic diffuse X-ray background radiation in the 15
-- 40 keV band, estimated to be 0.016 c s$^{-1}$. \label{fig2}}
\end{figure} 

\begin{figure}
\includegraphics[scale=0.5, angle=-90, clip] {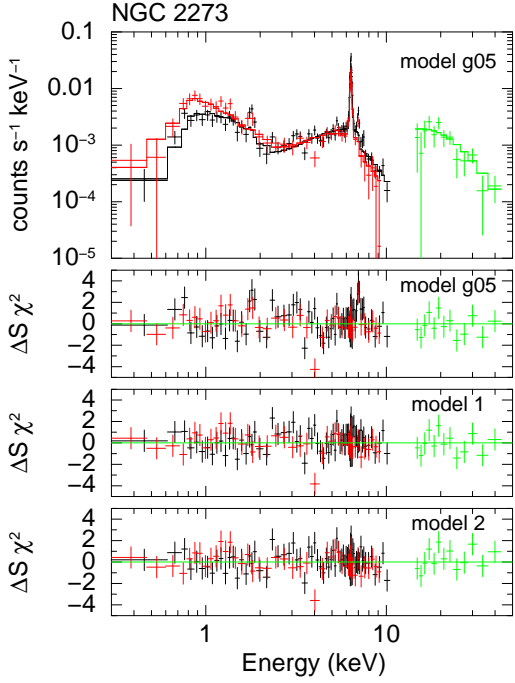}
\caption{A broad-band spectrum of NGC 2273 observed with Suzaku and
residuals for our fitting models.  The upper  panel shows the
wide-band spectrum obtained by XIS0+3 (black), XIS1 (red), and PIN
(green). The "model g05" corresponds to the model by
\citet{guainazzi05}. Models 1 and 2 are described in Table 2. }
\end{figure}


\begin{figure}
\includegraphics[scale=0.4, angle=-90, clip] {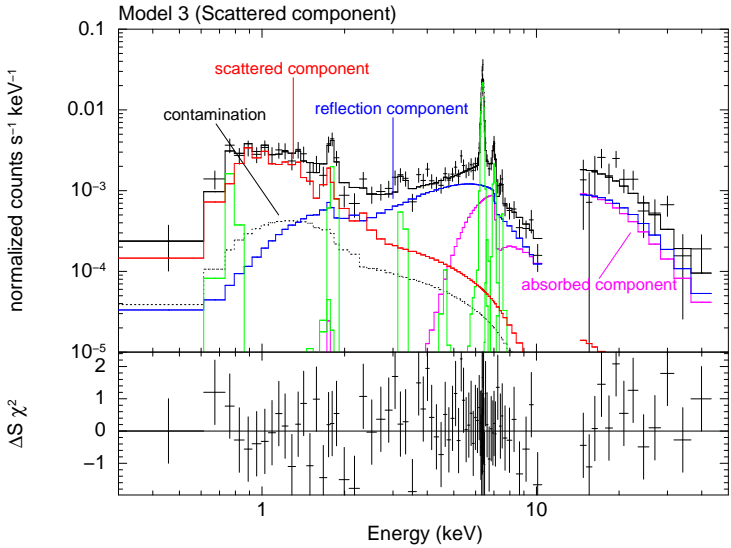}
\caption{The best-fit spectrum obtained by XIS0 and PIN for the model
3 in Table 2.   In this model, the soft component was fitted with a
scattered light. We also display the contamination (dashed line) from
the nearby bright source (see text). The green lines indicate emission
lines included in our fitting model. Escape lines for the emission
lines are also displayed in  this figure. 
}
\end{figure}

\begin{figure}
\includegraphics[scale=0.4, angle=-90, clip] {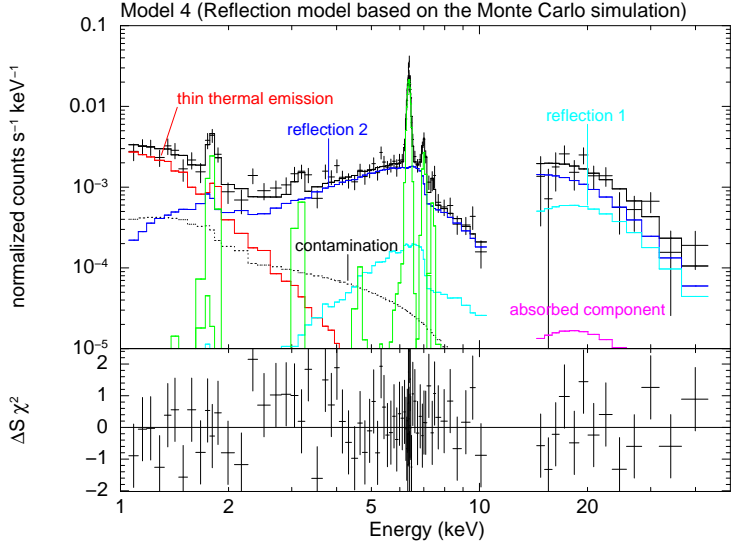}
\caption{The best-fit spectrum obtained by XIS0 and PIN for  model
4 in Table 4.   In this model, the reflection component was fitted
with the reflection model based on a Monte Carlo simulation (Ikeda et
al. 2008).  A brief explanation of reflections 1 and 2 is presented
in text. The green lines indicate emission lines included in our
fitting model. Escape lines for the emission lines are also displayed
in  this figure. The photon index of the power-law emission from the
central X-ray source  is fixed at 1.9. 
}
\end{figure}

\begin{figure}
\FigureFile(100mm,50mm){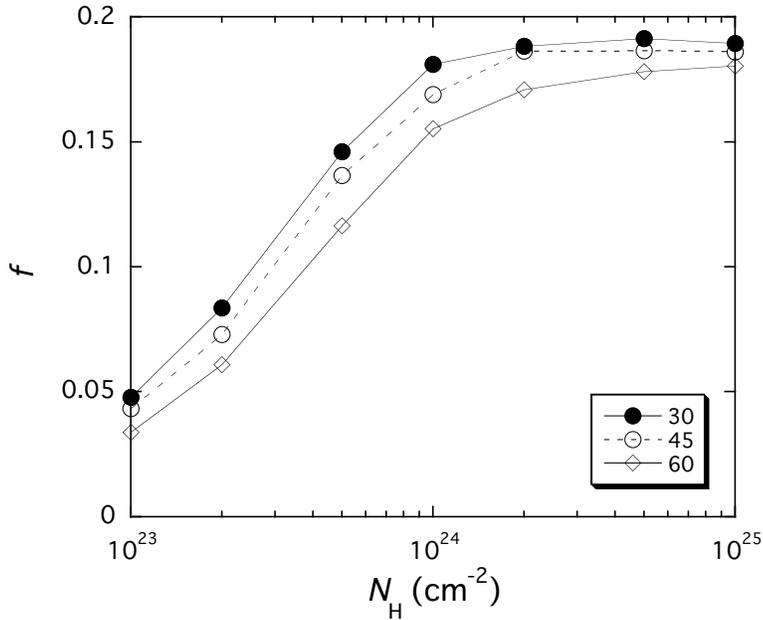}
\caption{Fractions of the Compton shoulder with respect to the primary
gaussian component as a function of the  column density along the
equatorial plane of a dusty torus. The fractions for the opening
angles of 30$^{\circ}$  (filled circles), 45$^{\circ}$ (open circles),
and 60$^{\circ}$ (open diamond) were estimated from the Monte Carlo
simulation developed by \citet{ikeda08}. The photon index of the
power-law emission from a central X-ray source  is fixed at 1.9. 
}
\end{figure}

\end{document}